\documentstyle[amsfonts,epsfig,12pt]{article}
\newcommand{\be}{\begin{equation}}
\newcommand{\ee}{\end{equation}}
\newcommand{\bea}{\begin{eqnarray}}
\newcommand{\eea}{\end{eqnarray}}
\newcommand{\beas}{\begin{eqnarray*}}
\newcommand{\eeas}{\end{eqnarray*}}
\newcommand{\bi}{\begin{itemize}}
\newcommand{\ei}{\end{itemize}}
\newcommand{\bc}{\begin{center}}
\newcommand{\ec}{\end{center}}
\newcommand{\bfl}{\begin{flushleft}}
\newcommand{\efl}{\end{flushleft}}
\newcommand{\bfr}{\begin{flushright}}
\newcommand{\efr}{\end{flushright}}

\def\6{\partial} \def\a{\alpha} \def\b{\beta}

\def\m{\mu} \def\n{\nu}  
\def\r{\rho} \def\s{\sigma}

\newcommand{\LL}{{\cal L}}

\begin{document}

\title{Bosonic D-branes at finite temperature with an external field}
\author{M. C. B. Abdalla, A. L. Gadelha, I. V. Vancea
\thanks{On leave from Babes-Bolyai University of Cluj.} \\
{\em Instituto de F\'{\i}sica Te\'{o}rica, Universidade Estadual
Paulista}\\
{\em Rua Pamplona 145, 01405-900, S\~{a}o Paulo, SP, Brasil}}
\maketitle

\begin{abstract}
Bosonic  boundary  states at  finite  temperature  are constructed  as
solutions of boundary conditions at $T\neq 0$ for bosonic open strings
with a  constant gauge  field $F_{ab}$ coupled  to the  boundary.  The
construction is  done in  the framework of  the thermo  field dynamics
where a  thermal Bogoliubov  transformation maps states  and operators
to  finite temperature.   Boundary states  are
given in  terms of  states from the  direct product space  between the
Fock space of  the closed string and another identical  copy of it. By
analogy  with   zero  temperature,   the  boundary  states   have  the
interpretation  of  $Dp$-branes at  finite  temperature. The  boundary
conditions admit  two different solutions.  The entropy of  the closed
string  in  a  $Dp$-brane  state  is  computed  and  analysed.  It  is
interpreted as the entropy of the $Dp$-brane at finite temperature.

\end{abstract}

\newpage

\section{Introduction}

There have been various reasons for which string theories at finite 
temperature have been considered an interesting subject \cite{lag}. 
In the past,
it was suggested in \cite{djg} that high-energy, fixed-angle scattering
could give some information about the physics of string theory as,
for example, the spontaneously broken symmetry group. The properties of an 
ideal gas of superstrings were used to model the cosmology of the early 
universe in \cite{ea,eao,yl1,bv,kp,akc}. The high-energy
behaviour of the thermal ensemble of the supersymmetric string theory showed 
that the underlying degrees of freedom of string theory is less than
it is known in relativistic field theories and that there is a first-order 
transition at the Hagedorn temperature in an ideal gas of strings \cite{aw}. 
To investigate the behaviour of string theory near the Hagedorn temperature,
a real-time finite-temperature interacting strings technique was developped in
\cite{yl2,yl3,eah}. 

On another line of development, it was proved that the $Dp$-branes belong to
the string spectra and couple with Ramond-Ramond fields \cite{jp}. The 
$Dp$-branes play a crucial role in understanding the string theory and its
connection with field theories and gravity \cite{jpb,cj}. Also, they have been
used to understand the statistical properties of various systems. 
The energy-entropy relation, the Hawking temperature and the 
Hagedorn transition of extreme, near-extreme and Schwarzschild black-holes
were calculated from the statistics of the Boltzman gas of free $D0$-branes
and from the counting of microstates in certain conformal field theories
\cite{as,sv,ls,hp,bfks1,cm,ks,eh,dmkr,hm,ml,bfk,lt,dm,ast,ivvol,ms}. However, 
despite the relative well knowledge of strings at finite temperature and the
positive results obtained from $Dp$-brane ensembles, not much is known about
the statistical properties of the $Dp$-branes (but see 
\cite{mvm,od,rsz,kog1,kog2,kog3,rab1,rab2,gub}).

In the absence of background fields, a $Dp$-brane  
has a microscopic description as a boundary 
state in the Fock space of the perturbative closed string where the 
brane is 
represented as a particular superposition of some coherent states on the
boundary of the cylinder. This description continues to hold when a
a gauge field is turned on
\cite{clny,pc,gw,bg,gg,ml1,ck,dv1,dv2}.
Since the
$Dp$-brane is written in terms of string operators acting on the vacuum,
one can apply the thermo field method \cite{tu} to construct the boundary
states at finite temperature. Since the states should belong to the
physical spectrum of the string, the contribution of ghosts should be
taken into account \cite{cav}. The thermo field dynamics was used previously
to study the renormalization of open bosonic strings at finite temperature
within the framework. The renormalizability was proved and the model 
was showed to be compatible with the thermal Veneziano amplitude
in \cite{fn1,hf}. The global phase structure of bosonic-thermal-string 
ensemble and its connection with the thermal string amplitude was 
described in \cite{fn2}.

In a previous paper \cite{ivv}, the Bogoliubov transformations of thermo field
were used to construct the bosonic string at finite temperature in a flat 
Minkowski
space-time and in the light-cone gauge. The boundary conditions necessary to
define the $Dp$-branes at finite temperature as states in the Fock space were 
obtained and it was shown that the corresponding equations admit solutions 
that can be interpreted as bosonic branes. Some particular solutions that 
reduce to the known $Dp$-branes at $T=0$ were also given. However, the analysis
was not complete.

The aim of this paper is to report on some new solutions of the bosonic 
boundary conditions at $T \neq 0$ and generalize the construction to a 
nontrivial background in which an $SO(1,p)$ photon $A^{a}$ is present on
the world-volume of the brane. To keep the program simple, we consider that
the photon does not depend on the temperature, although the most general case
can be written down by recalling that the photon belongs to the spectrum
of the open strings that ends on the brane. We also give the solutions of the
boundary conditions which are a superposition of closed string states at 
finite temperature. By analogy with $T=0$ we interpret the boundary states
as $Dp$-branes at $T \neq 0$. Within the framework of thermo field, it is 
possible to compute the entropy as the expectation value of the entropy
operator. 
We compute the value of this entropy in a $Dp$-brane state and we show that
it has the following behaviour. In the limit of
low temperatures $T \rightarrow 0$, the contribution of the closed string 
oscillators diverge and one is left with a  series in the powers of the
$({\Bbb I}-{\Bbb F}/{\Bbb I}+{\Bbb F})$ where $F$ is the gauge field. However,
in the limit when $T \rightarrow \infty$ the oscillator contribution is 
proportional to $ \log(-1)$ which does not make sense. This might be a
consequence of the fact that above a critical temperature (the Hagedorn
temperature) the notion of temperature breaks down in string theory. If this
is the case, even if formally we can construct the boundary states at any 
temperature, they do not make sense at temperatures where the string model 
ceases to make sense.
We do not have a good explanation of the general behaviour of the entropy. 
It is premature
to say, even if the construction of these states is 
acceptable, if the behaviour of the entropy signals that the bosonic 
$Dp$-branes cannot be described as acceptable states in the Fock space 
of closed bosonic string theory. However, before drawing such of conclusions,
deeper study of the model should be performed and we refere the reader to 
\cite{cawv} for further discussions of these problems.  

The paper is organized as follows. In Section 2 we construct
the closed string at finite temperature in the thermo field approach and 
give the boundary conditions and the $Dp$-brane at $T \neq 0$. In Section 3 we 
compute the entropy of $Dp$-brane states. Section 4 is devoted to
discussions. The basic relations of thermo field
dynamics which have been used throughout the text are presented in the 
Appendix.  

\section{Boundary states at finite temperature with an external field}

In this section we construct the boundary states of the bosonic closed
strings in the presence of an open string $SO(1,p)$ vector field at finite
temperature. We then give the solutions of the corresponding equations. 

\subsection{Closed strings at $T \neq 0$ }

Consider the bosonic open string on flat Minkowski space-time in
the presence of a bosonic and rigid $Dp$-brane located at the boundary 
$\s=0$ of the world-sheet. The $U(1)$ 
charges at the endpoints of the string generate
a vector field $A^{a}$, that lives on the world-volume of the brane.
To avoid dealing with ghosts, we choose to work in the light-cone gauge
$X^0 \pm X^{25}$.
Then $a,b,... =1,\ldots,p$ denote the indices of the tangential coordinates
to the world-volume. For the transversal coordinate we use the indices
$i,j,... = p+1, \ldots,24$. The space-time indices are labelled by 
$\m, \n ,...= 1,\ldots,24$. By interpreting the one-loop diagram in open
string amplitude as the tree-level diagram in the closed string channel,
one can give a microscopic description of the $Dp$-brane as a state in
the Fock space of the free closed string. This state is defined by $p+1$
Neumann boundary conditions and $25-p$ Dirichlet boundary conditions acting
on a vector of the Fock space.
At $T=0$ and in the presence of the $SO(1,p)$ photon the 
corresponding relations, written in terms of string modes, are given by
\begin{eqnarray}
\left[ \left( {\Bbb I}+\widehat{{\Bbb F}}\right) _{b}^{a}\alpha
_{n}^{b}+\left( {\Bbb I}-\widehat{{\Bbb F}}\right) _{b}^{a}\beta
_{-n}^{b} \right]\left| B\right\rangle & = & 0
\nonumber\\
\left( \alpha _{n}^{i}-\beta _{-n}^{i}\right) \left| B\right\rangle &=& 0
\label{boundcond1}\\
\hat{p}^{a}\left| B\right\rangle =\left(\hat{q}^{i}-y^{i}\right) \left|
B\right\rangle & = & 0,\nonumber
\end{eqnarray}
for any $ n > 0 $, where $\a^{a}_n$ and $\b^{b}_n$ are respectively the
right- and left-moving modes of the
closed string.  $q^{i}$ and $\hat{p}^{a}$ are components of the 
coordinate of the center of mass and its momenta, respectively, and $y^i$ are
the coordinates of the $Dp$-brane in the transverse space. We use for the
scaled gauge field the notation
\begin{equation}
\widehat{{\Bbb F}}_{ab}=2\pi \alpha ^{\prime }F_{ab},
\label{scaledgauge}
\end{equation}
where $ F_{ab}$ is constant. The state that satisfies the boundary conditions 
(\ref{boundcond1}) is denoted by $ \left| B \right\rangle$. This state belongs
to the bosonic Fock space. In a covariant gauge one should take into account
the ghost contribution then the brane state factorizes in a term that satisfies
(\ref{boundcond1}) and another one that contains the ghost contribution. The
solution to (\ref{boundcond1}) is a superposition of coherent states of the
following form \cite{dv1,dv2}
\begin{eqnarray}
\left| B\right\rangle &=&\exp \left[ -\sum\limits_{n=1}^{\infty }\alpha
_{-n}^{\mu }M_{\mu \nu }\beta _{-n}^{\nu }\right] \left| B\right\rangle
^{\left( 0\right) } \\
\left| B\right\rangle ^{\left( 0\right) } &=&N_{p}\delta ^
{\left(d_{\perp }\right) }\left( \hat{q}-y\right) \left| 0\right\rangle ,
\label{soltempzero}
\end{eqnarray}
where $N_{p}$ is a normalization constant equal to half of the
brane tension and the delta function localizes the state in the trasverse 
space in the position $q^i$. The vacuum state includes a term which is 
an eigenvector of the momentum operators which is not written explicitly.
The matrix that connects the right and left modes is given by the following 
relation
\begin{equation}
M_{\nu }^{\mu }=\left[ \left( \frac{{\Bbb I}-{\Bbb F}}{{\Bbb I}+{\Bbb F}}
\right) _{b}^{a};-\delta _{j}^{i}\right].
\end{equation}

Our first purpose is to construct the finite temperature counterpart of
the boundary states (\ref{boundcond1}). Since the relations are expressed in 
terms of bosonic string operators and states, one has to map these objects
at $T \neq 0$. A convenient way to do that is by employing thermo field
techniques \cite{tu} which are suitable for systems that are represented in
terms of oscillators. (Some basic relations of this construction are presented
in Appendix A.) The thermo field approach was used previously to study strings 
in \cite{fn1,hf,fn2,cav} and $Dp$-branes in \cite{ivv}. According to the
thermo field dynamics, the thermodynamics of the system is described in an
enlarged Fock space (which is also true in the path integral approach) 
composed by the initial Fock space and an identical copy of it. This is what
is meant by ``doubling the system''. The total thermic system is composed
by the original string and its copy, denoted by $\widetilde{}$ . The
two copies are independent and the Fock space is the direct product of them.

In order to implement this construction for the case of the bosonic string 
\cite{ivv}, we 
use the oscillator operators for left and right modes
\begin{eqnarray}
A_{n}^{\mu } &=&\frac{1}{\sqrt{n}}\alpha _{n}^{\mu },\quad A_{n}^{\mu
\dagger }=\frac{1}{\sqrt{n}}\alpha _{-n}^{\mu }, \\
B_{n}^{\mu } &=&\frac{1}{\sqrt{n}}\beta _{n}^{\mu },\quad B_{n}^{\mu \dagger
}=\frac{1}{\sqrt{n}}\beta _{-n}^{\mu },
\label{oscillators}
\end{eqnarray}
where $n>0$. Identical operators exist for the copy of the system. The
two algebras are independent
\be
[ A^{\mu}_{n}, A^{\nu\dagger}_{m}] = [{\tilde{A}}^{\mu}_{n},
{\tilde{A}}^{\nu\dagger}_{m}] = \delta_{n,m}\eta^{\mu \nu},
\ee
\be
[A^{\mu}_{n},{\tilde{A}}^{\nu}_{m}] = 
[A^{\mu}_{n},{\tilde{A}}^{\nu\dagger}_{m}] 
= [A^{\mu}_{n},{\tilde{B}}^{\nu}_{m}]=\cdots = 0.
\label{comm}
\ee
The extended Fock space of the total system is given by the direct product 
of the two Fock spaces of closed strings
\begin{equation}
\widehat{H}=H\otimes \widetilde{H}.
\label{fock}
\end{equation}
A state from $ \widehat{H}$ is denoted  by $|~~\rangle \rangle$. 
The vacuum states of the left and right sectors are direct products
of the vacuum states of the string and tilde-string 
\begin{eqnarray}
\left. \left| 0\right\rangle \! \right\rangle _{\alpha } &=&\left|
0\right\rangle _{\alpha }\otimes \widetilde{\left| 0\right\rangle }_{\alpha
}=\left| 0,0\right\rangle _{\alpha }, \\
\left. \left| 0\right\rangle \! \right\rangle _{\beta } &=&\left|
0\right\rangle _{\beta }\otimes \widetilde{\left| 0\right\rangle }_{\beta
}=\left| 0,0\right\rangle _{\beta },
\label{vacuum}
\end{eqnarray}
and the following expressions are equal among them
\bea
|0\rangle \! \rangle &=& |0 \rangle \! \rangle_{\a} |0 \rangle \! \rangle_{\b} 
= (|0\rangle_{\a} |\tilde{0}\rangle_{\a})
(|0\rangle_{\b}|\tilde{0}\rangle_{\b})
\nonumber\\
&=&(|0\rangle_{\a}|0\rangle_{\beta })
(|\tilde{0}\rangle_{\a}|\tilde{0}\rangle_{\beta }),
\label{vactzero1}
\eea
where the last equality is a consequence of the fact that the original string 
and the tilde-string are independent. The first line in (\ref{vactzero1}) shows
explicitly the doubling of each oscillator while the second one shows the
string-tilde-string structure of the vacuum state. In order to obtain
the fundamental state of the enlarged system we have to multiply
(\ref{vactzero1}) by $|p \rangle |\tilde{p} \rangle$. 

The extended system represents the general framework for studing the thermal
properties of the initial one. The finite temperature is introduced by a 
set of Bogoliubov unitary operators acting on the states as well as on the
operators of the extended system. These operators are constructed for each of 
the oscillating modes of the two copies and according to the thermo field 
construction their form in the right and left sectors are given by the 
following relations
\begin{equation}
G_{n}^{\alpha }=-i\theta \left( \beta_T \right) \left( A_{n}\cdot
\widetilde{A}
_{n}-A_{n}^{\dagger }\cdot \widetilde{A}_{n}^{\dagger }\right), 
\label{bogolibopright}
\end{equation}
\begin{equation}
G_{n}^{\beta }=-i\theta \left( \beta_T \right) \left( B_{n}\cdot \widetilde{B}%
_{n}-B_{n}^{\dagger }\cdot \widetilde{B}_{n}^{\dagger }\right). 
\label{bogolibopleft}
\end{equation}
Here, $\beta_T = (k_BT)^{-1}$ where $k_B$ is the Boltzmann's constant and 
$\theta$ is a parameter depending on the temperature hyperbolically as
\begin{equation}
\cosh \theta _{n}\left( \beta_{T} \right) = u_{n}\left( \beta_{T} \right), 
\label{theta}
\end{equation}
where $u_{n}\left( \beta_{T} \right)$ is related to the statistics of the
modes. For bosonic oscillators two hyperbolic trigonometric functions are 
defined (see the Appendix)
\begin{eqnarray}
u_{n}\left( \beta_{T} \right)  &=&\left( 1-e^{-\beta_{T} \omega _{n}}\right)
^{-1/2}, \\
v_{n}\left( \beta_{T} \right)  &=&\left( e^{\beta_{T} \omega _{n}}-1\right)
^{-1/2}.
\end{eqnarray}

Let us look at the properties of the Bogoliubov operators.
From the definition (\ref{bogolibopright}) and (\ref{bogolibopleft}) 
it is easy to see that they are Hermitian  
\begin{equation}
\left( G_{n}^{\alpha }\right) ^{\dagger }=G_{n}^{\alpha },\quad \left(
G_{n}^{\beta }\right) ^{\dagger }=G_{n}^{\beta },
\label{hermgop}
\end{equation}
and that for negative $n$ the following relation holds
\begin{equation}
G_{\left| n\right| }^{\alpha }=-G_{-n}^{\alpha }.
\label{negativenop}
\end{equation}
Since the right- and left-moving modes are independent, the corresponding 
$G$-operators commute among themselves
\begin{equation}
\left[ G_{n}^{\alpha },G_{m}^{\alpha }\right] =\left[ G_{n}^{\beta
},G_{m}^{\beta }\right] =\left[ G_{n}^{\alpha },G_{m}^{\beta }\right] =0.
\label{algebraG}
\end{equation}
A simple algebra gives the rest of the commutation relations among the
$G$-operators and the oscillators
\begin{eqnarray}
\left[ G_{n}^{\alpha },A_{n}^{\mu }\right] &=&-i\theta _{n}\left( \beta_{T}
\right) \widetilde{A}_{n}^{\mu \dagger },\qquad \left[ G_{n}^{\alpha
},B_{n}^{\mu }\right] =-i\theta _{n}\left( \beta_{T} \right) \widetilde{B}
_{n}^{\mu \dagger },\nonumber\\
\left[ G_{n}^{\alpha },A_{n}^{\mu \dagger }\right] &=&-i\theta _{n}\left(
\beta_{T} \right) \widetilde{A}_{n}^{\mu },\qquad \left[ G_{n}^{\alpha
},B_{n}^{\mu \dagger }\right] =-i\theta _{n}\left( \beta_{T} \right)
\widetilde{B}_{n}^{\mu },\nonumber\\
\left[ G_{n}^{\alpha },\widetilde{A}_{n}^{\mu }\right] &=&-i\theta _{n}\left(
\beta_{T} \right) A_{n}^{\mu \dagger },\qquad \left[ G_{n}^{\alpha },
\widetilde{B}_{n}^{\mu }\right] =-i\theta _{n}\left( \beta_{T} \right)
B_{n}^{\mu \dagger },\nonumber\\
\left[ G_{n}^{\alpha },\widetilde{A}_{n}^{\mu \dagger }\right] &=&-i\theta
_{n} \left( \beta_{T} \right)A_{n}^{\mu },\qquad \left[ G_{n}^{\alpha },
\widetilde{B}_{n}^{\mu \dagger }\right] =-i\theta _{n}\left( \beta_{T} \right)
B_{n}^{\mu }.
\label{gopalgebra}
\end{eqnarray}

Let us proceed to the construction of the vacuum state and creation and
annihilation operators at $T \neq 0 $ following \cite{tu}. By acting on
the right and left vacuua at $T=0$ with any of the Bogoliubov operators 
(\ref{bogolibopright}) and (\ref{bogolibopleft}) new states, which depend 
explicitly on the temperature, are obtained
\bea
|0(\b_T ) \rangle \! \rangle_\a & = & \prod_{n>0}e^{-iG^{\a}_{n}}
|0\rangle \! \rangle_{\a} = 
\prod_{n>0}|0(\b_T )_n \rangle \! \rangle_\a, \\
|0(\b_T ) \rangle \! \rangle_\b & = &
\prod_{m>0}e^{-iG^{\b}_{m}}|0\rangle \! \rangle_{\b} =
\prod_{n>0}|0(\b_T )_n \rangle \! \rangle_\b.
\label{vactnzerolefrig}
\eea
Since the $G$-operators do not mix the left and right moving states, one
can construct a direct product of the states above
\be
|0(\b_T ) \rangle \! \rangle =
|0(\b_T ) \rangle \! \rangle_\a |0(\b_T ) \rangle \! \rangle_\b .
\label{vacuumatT}
\ee
The  Bogoliubov transformations acting on the oscillator operators
$ \{ A^{\dagger}, A, {\tilde{A}}^{\dagger}, \tilde{A} \}$ and 
$ \{ B^{\dagger}, B, {\tilde{B}}^{\dagger}, \tilde{B} \}$ map them
to some new operators that depend on temperature
\begin{eqnarray}
A_{n}^{\mu }\left( \beta_{T} \right) & = &e^{-iG_{n}^{\alpha }}A_{n}^{\mu
}e^{iG_{n}^{\alpha }}~~,~~
B_{n}^{\mu }\left( \beta_{T} \right) =  e^{-iG_{n}^{\alpha }}B_{n}^{\mu
}e^{iG_{n}^{\alpha }}, \\
\widetilde{A}_{n}^{\mu }\left( \beta_{T} \right) & = & e^{-iG_{n}^{\alpha }}
\widetilde{A}_{n}^{\mu }e^{iG_{n}^{\alpha }}~~,~~
\widetilde{B}_{n}^{\mu }\left( \beta_{T} \right) =  e^{-iG_{n}^{\alpha }}
\widetilde{B}_{n}^{\mu }e^{iG_{n}^{\alpha }}.
\label{finitetimpop1}
\end{eqnarray}
From the properties (\ref{hermgop})-(\ref{gopalgebra}) one can see that the
oscillator operators at finite temperature can be cast in the following form
\bea
A_{n}^{\mu }\left( \beta_{T} \right) & = &
u_{n}\left( \beta_{T} \right) A_{n}^{\mu }-v_{n}\left(
\beta \right) \widetilde{A}_{n}^{\mu \dagger}, \nonumber \\ 
B_{n}^{\mu }\left( \beta_{T} \right) & = &
u_{n}\left( \beta_{T} \right) B_{n}^{\mu }-v_{n}\left(
\beta \right) \widetilde{B}_{n}^{\mu \dagger }, \nonumber\\
\widetilde{A}_{n}^{\mu }\left( \beta_{T} \right) & = & 
u_{n}\left( \beta_{T} \right) 
\widetilde{A}_{n}^{\mu }-v_{n}\left( \beta_{T} \right)
A_{n}^{\mu \dagger }, \nonumber \\
\widetilde{B}_{n}^{\mu }\left( \beta_{T} \right) & = &
u_{n}\left( \beta_{T} \right) 
\widetilde{B}_{n}^{\mu }-v_{n}\left( \beta_{T} \right) B_{n}^{\mu \dagger }
\label{finitetempop2}.
\eea
The notation for the state (\ref{vacuumatT}) is now justified. Indeed, this 
state is annihilated by the finite temperature operators in both left and right
sectors and for the string and its copy 
\bea
A_{n}^{\mu }\left( \beta_{T} \right) \left. \left| 0\left( \beta_{T} \right)
\right\rangle \! \right\rangle _{\alpha } & = &\widetilde{A}_{n}^{\mu }
\left( \beta
\right) \left. \left| 0\left( \beta_{T} \right) \right\rangle \! \right\rangle
_{\alpha }=0, \\
B_{n}^{\mu }\left( \beta_{T} \right) \left. \left| 0\left( \beta_{T} \right)
\right\rangle \! \right\rangle _{\beta }& = &\widetilde{B}_{n}^{\mu }\left( \beta
\right) \left. \left| 0\left( \beta_{T} \right) \right\rangle \! \right\rangle
_{\beta }=0,
\label{vacuumannih}
\end{eqnarray}
which shows that (\ref{vacuumatT}) has the properties of the vacuum state.

The finite temperature operators statisfy the oscillator algebra for each mode,
in each sector and for both copies of the original string and all these
algebras are independent. Therefore, the states of the system at finite 
temperature are obtained by acting on the thermal vacuum (\ref{vacuumatT})
with the creation and annihilation operators. These states belong to the total
Fock space. The Bogoliubov transformations have picked up another vacuum and 
other oscillators. However, we should note that the thermal vacuum contains
infinitely many quanta of zero temperature operators of all types due to the 
form of the Bogoliubov operators. By creating a quanta of, say, $A$
type, one of $\widetilde{A}$ type is destroyed which suggests that the
second copy of the system be interpreted as a thermal reservatoire \cite{tu}.  

An important question is whether the thermo field construction presented
above gives a string theory at $ T \neq 0$ or maps it into something else.
If we plug the finite temperature operators into a solution of the equations
of motion of closed bosonic string, one obtains another solution which
depends on $T$ 
but which {\em mixes the original string and the tilde-string} in an obvious
way.\footnote{This results from the definition of $G$-operators which do not
affect the world-sheet waves but only the Fourier coefficients. The new
Fourier coefficients determine a new solution.} One can show that all the
properties of the bosonic string at zero temperature are satisfied.
In particular, one can construct from the solutions of the equations of
motion, the energy-momentum tensor that has the same form as the one
at $T=0$. Then the following operators
\begin{equation}
L_{m}^{\alpha }\left( \beta_{T} \right) =\frac{1}{2}\sum\limits_{k\in {\Bbb Z}
}\alpha _{-k}\left( \beta_{T} \right) \cdot \alpha _{k+m}\left( \beta_{T} \right), 
\label{Virasoroop}
\end{equation}
satisfy the Virasoro algebra as can be shown by using the properties of 
the Bogoliubov operators (\ref{hermgop})-(\ref{gopalgebra}). Therefore, the 
conformal symmetry is not broken by the present constructionat least on-shell.
Since we are dealing with two copies of the same system, all that was said
above is true for the tilde-string. However, note that at finite temperature
the notions of string at tilde-string are slightly different since both of
them mix the operators of string and tilde-string as they were defined
initially at $T=0$. In this sense tilde symbol just reminds us that
the operators were obtained from the copy of the original system.

\subsection{Boundary conditions at $T\neq 0$}

To define the $Dp$-branes at $T\neq 0$ we have to construct the counterpart
of the boundary conditions (\ref{boundcond1}) at finite temperature for
both, the string and the tilde-string. We have three possibilities to do that:
either to map the operators that describe the relations (\ref{boundcond1})
at finite $T \neq 0$, to map the operators and the states or 
to map the states at finite temperature while keeping the operators at 
$T = 0$. The first two alternatives give the solutions at finite 
temperature. The last one gives no new information since its solutions
should be among the solutions at zero temperature.  This reminds the
different pictures in quantum mechanics and it would be interesting to see
the relation between all these representations. In what follows we will map
the operators at $T=0$ using the Bogoliubov transformation and it will turn
out that the states that satisfy these boundary conditions depend explicitly 
on $T$ \cite{ivv}. It is possible to derive formally the boundary 
conditions from an action.
Since the properties of the bosonic
string are satisfied, the string equations can be obtained from a 
formal string action which mimics the zero temperature action and depends 
implicitly on the parameter $\b_T$. Its form is given by the following 
relation
\begin{eqnarray}
S\left( \beta_{T} \right)  &=&\frac{1}{4\pi \alpha ^{\prime }}\int d\tau \int
d\sigma \left\{ \partial ^{\alpha }X^{\mu }\left( \beta_{T} \right) \partial
_{\alpha }X_{\mu }\left( \beta_{T} \right) \right. +  \nonumber \\
&&-\left. \left[ \delta \left( \sigma \right)  \dot{X}^{\mu }
\left( \beta_{T} \right) A_{\mu }\left[ X\left( \beta_{T} \right) \right]
\right] \right\}, 
\label{actionT}
\end{eqnarray}
which is the action of a open string that couples with an $U(1)$ field on 
a $Dp$-brane located at $\s =0$ on the world-sheet. We assume that the gauge 
field is nonzero along the directions parallel to the world-volume of the 
brane and 
$A^i(X(\b_T )) = \mbox{constant}$. Moreover, we do not consider any explicit
dependence on the temperature of the gauge field. Then the boundary conditions 
in the closed string sector take the usual form  
\begin{eqnarray}
\left. \left( \partial _{\tau }X_{a}\left( \beta_T \right) +F_{ba}\partial
_{\sigma }X^{b}\left( \beta_T \right) \right) \right| _{\tau =0} &=&0,\\
\left. X^{i}\left( \beta_T \right)-y^{i} \right| _{\tau =0}&=&0.
\label{boundT}
\end{eqnarray}
Since the solutions of the equations of motion that enter (\ref{boundT}) have
the form of a closed string solution with the oscillator operators replaced by
the operators at $T\neq 0$, one can express (\ref{boundT}) in terms of creation
and annihilation operators  
\begin{eqnarray}
\left( {\Bbb I}+{\Bbb F}\right) _{b}^{a}A_{n}^{b}\left( \beta_{T} \right)
+\left( {\Bbb I}-{\Bbb F}\right) _{b}^{a}B_{n}^{b\dagger }\left( \beta_{T}
\right)  &=&0, \nonumber \\
\left( {\Bbb I}+{\Bbb F}\right) _{b}^{a}A_{n}^{b\dagger }\left( \beta_{T}
\right) +\left( {\Bbb I}-{\Bbb F}\right) _{b}^{a}B_{n}^{b}\left( \beta_{T}
\right)  &=&0, \nonumber \\
A_{n}^{i}\left( \beta_{T} \right) -B_{n}^{i\dagger }\left( \beta_{T} \right) 
&=&0, \nonumber \\
A_{n}^{i\dagger }\left( \beta_{T} \right) -B_{n}^{i}\left( \beta_{T} \right) 
&=&0,
\label{boundarycondT}
\eea
for any $n>0$ which are supplement by the boundary conditions for the
position of center of mass of string and its conjugate momenta
\bea
p^{a} &=&0,\\
q^{i}-y^{i} &=&0.
\label{condmomT}
\end{eqnarray}
The last two relations are a consequence of the fact that 
the operators $\hat{p}$, $\hat{X}$, $\hat{\tilde{p}}$ and 
$\hat{\tilde{X}}$ commute with all oscillator operators. Therefore, they are 
not affected by the Bogoliubov transformations. If
we construct the zero mode $G$-operators as we did for the oscillators
we see that the momenta commute with them. Thus, we can take the position and 
momenta operators of both, the string and the tilde-string, to be invariant
under the transformations above. Also the corresponding eigenstates of the
momenta operators are taken to be invariant.  

Some comments are in order now. We note that the above construction should
be applied to the tilde-string as well. Thus, we will have two boundary
conditions, where the second one is identical to (\ref{boundarycondT}) with
all the operators replaced by the corresponding tilde operators.
These boundary conditions can be constructed with the same mapping or from
a formal action of the form (\ref{actionT}). It is not possible to construct
the boundary conditions from a Lagrangian at finite temperature since that is
not the basic object in thermo field dynamics which deals only with a Lagrangian
which depends on the fields at $T=0$ of the following form
$ \LL - \widetilde{\LL}$ where the first term contains information about the 
string and the second term is the Lagrangian of the tilde-string \cite{tu}. 
Then the boundary conditions should be imposed as we have already commented
at the beginning of this paragraph, i.e. by mapping them to finite temperature.
We could have started from the thermo field Lagrangian to obtain the boundary
conditions at $T=0$, which are just our conditions since the two terms in the
thermo field Lagrangian are 
independent and then map each of them at $ T \neq 0$. 
Therefore, solving the
boundary conditions within the thermo field framework is equivalent to
solving them directly from the Lagrangian (\ref{actionT}) and its copy.
 
\subsection{Boundary states at $T\neq 0$}

Our next goal is to solve the equations (\ref{boundarycondT}) and the 
corresponding ones for the tilde-fields. To this end, we note that
(\ref{boundarycondT}) have the same form as (\ref{boundcond1}).
Moreover, the operators at $T \neq 0 $ and the thermal
vacuum have the same properties as the ones at $ T = 0 $. Therefore, the 
solutions to the boundary conditions (\ref{boundarycondT}) is given by
\be
\left. \left| B\left( \beta_T \right) \right\rangle \! \right\rangle 
={N}_{p}\left( F,\beta_T \right) \delta ^{\left( d_{\perp }\right) }\left( 
\hat{q}-y\right)
e^{-\sum\limits_{n=1}^{\infty }A^{\mu \dagger }
\left( \beta_T \right) 
M_{\mu \nu }B^{\nu \dagger }\left( \beta_T \right)} \left.
\left| 0\left( \beta_T \right) \right\rangle \! \right\rangle ,
\label{solT11}
\ee
where $N_p\left( F, \b_T \right)$ is the normalization constant at $T \neq 0$.
An eigenstate of the momentum of the center of mass is included in the 
thermal vacuum.
In a similar manner one can write the solution to the other boundary condition
for the tilde-string. The total solution will be a product of them, and it is 
given by the following formula
\bea
\left. \left| B\left( \beta_T \right) \right\rangle \! \right\rangle _{1} 
&=&N_{p}^{2}\left( F,\beta_T \right) \delta ^{\left( d_{\perp}\right) }
\left( \hat{q}-y\right) \delta ^{\left( d_{\perp }\right) }\left( 
\widetilde{\hat{q}}-\widetilde{y}\right) \times \nonumber\\   
&&e^{-\sum\limits_{n=1}^{\infty }
A_{n}^{\dagger }
\left( \beta_T \right) 
\cdot M\cdot
B_{n}^{\dagger }
\left( \beta_T \right)}
e^{-\sum\limits_{n=1}^{\infty }
\widetilde{A}_{n}^{\dagger }
\left( \beta_T \right) 
\cdot M\cdot 
\widetilde{B}_{n}^{\dagger }
\left( \beta_T \right) } \left. 
\left| 0\left( \beta_T \right) \right\rangle \! \right\rangle.
\label{solT1}
\eea
Here, the normalization constants for the two solutions were taken equal
because we deal with identical copies of the system. To
compute it we should calculate the scattering amplitude in the closed string
channel and in the open string channel and compare the results. However, since 
the solutions are formally identical to the ones at $T=0$ one can set 
$ N_{p}\left( F,\beta_T \right) =  N_{p}\left( F \right)$ which is known to 
be proportional to the Born-Infeld action \cite{dv1,dv2}
\be
N_{p}\left( F,\beta_T \right) = N_p \left( F \right) = 
\sqrt{- \mbox{det}(\delta + \hat{F})}.
\label{normconst}
\ee
The solution (\ref{solT1}) was obtained using just the algebra of the operators
$A_{\b_T}$, $B_{\b_T}$, $\widetilde{A}_{\b_T}$, $\widetilde{B}_{\b_T}$ 
and $G$-operators. Due to their action on the vacuum be similar to those at
$T=0$, (\ref{solT1}) has formally the same form as the boundary
state at zero temperature (\ref{soltempzero}). The difference is that the
new solutions contain an explicit dependence on the temperature.  

Another solution can be obtained observing that the boundary condition 
(\ref{boundarycondT}) can be written as
\begin{eqnarray}
\left( {\Bbb I}+{\Bbb F}\right) _{b}^{a} 
e^{-iG^{\a}_n} A_{n}^{b} e^{iG^{\a}_n}
+\left( {\Bbb I}-{\Bbb F}\right) _{b}^{a}
e^{-iG^{\b}_n} B_{n}^{b\dagger } e^{iG^{\b}_n}
&=&0, \nonumber \\
\left( {\Bbb I}+{\Bbb F} \right) _{b}^{a}
e^{-iG^{\a}_{n}} A_{n}^{\dagger b}
e^{iG^{\a}_n} +\left( {\Bbb I}-{\Bbb F} \right) _{b}^{a}
e^{-iG^{\b}_n} B_{n}^{b} e^{iG^{\b}_n}
&=&0, \nonumber\\
e^{-iG^{\a}_n} A_{n}^{i} e^{iG^{\a}_n}- 
e^{-iG^{\b}_n} B_{n}^{i\dagger } e^{iG^{\b}_n}
&=&0, \nonumber \\
e^{-iG^{\a}_{n}} A_{n}^{i\dagger } e^{iG^{\a}_n}+
e^{-iG^{\b}_n} B_{n}^{b} e^{iG^{\b}_n}
&=&0,
\label{boundcondTgop}
\eea
for any $n>0$. Similar equations can be written for the boundary of the
tilde-string with tilde operators instead. Then one can
see that there is a solution to the equations (\ref{boundcondTgop}) of the
form
\be
\left| B (\b_T) \right\rangle \! \rangle _2 = N_p(\b_T )
\prod_{n=1}^{\infty }e^{-iG^{\a}_{n}} \prod_{m=1}^{\infty } e^{-iG^{\b}_m}
\left| B\right\rangle,
\label{solT21}
\ee
where $\left| B\right\rangle$ is the boundary state  at $T=0$ given in
(\ref{soltempzero}). A similar expression can be found for the tilde operators
and the second solution to the boundary conditions at $T \neq 0$ has the form
\be
\left| B (\b_T) \right\rangle \! \rangle _2 =
\left| B\right\rangle \widetilde{\left| B (\b_T) \right\rangle }.
\label{solT2}
\ee   
Note that the solutions (\ref{solT1}) and (\ref{solT2}) are different because
the exponential contains different operators. They have a natural degeneracy 
due to the fact that we work in a doubled Fock space and are physical
solutions in the light-cone gauge. In a conformal gauge, the contributions
of ghosts at finite temperature \cite{cav} should be taken into account in
order to eliminate the unphysical degrees of freedom of closed string. In
what follows we study the solution (\ref{solT1}). Solution
(\ref{solT2})
represents the mapping of the boundary state at $T=0$ at finite temperature
via Bogoliubov operators. 

\section{Entropy of D-branes}

It is interesting to investigate now the thermal properties of the boundary 
states obtained in the previous section. Since we are working with closed
string, we are going to compute its entropy in the boundary state 
(\ref{solT1}). 

According to the thermo field dynamics \cite{tu}, the entropy
operators for the bosonic closed string in terms of string oscillators
and in $k_B$ units are given by the following relations 
\begin{eqnarray}
K &=&\sum\limits_{\mu }\sum\limits_{n}\left[ \left( A_{n}^{\mu \dagger
}A_{n}^{\mu }+B_{n}^{\mu \dagger }B_{n}^{\mu }\right) \log \sinh ^{2}\theta
_{n}\right. +  \nonumber \\
&&\left. -\left( A_{n}^{\mu }A_{n}^{\mu \dagger }+B_{n}^{\mu }B_{n}^{\mu
\dagger }\right) \log \cosh ^{2}\theta _{n}\right], 
\label{entroop1}
\end{eqnarray}
for the string, and 
\begin{eqnarray}
\widetilde{K} &=&\sum\limits_{\mu }\sum\limits_{n}\left[ 
\left( \widetilde{A}_{n}^{\mu
\dagger }\widetilde{A}_{n}^{\mu }+\widetilde{B}_{n}^{\mu \dagger }
\widetilde{B}_{n}^{\mu }\right) \log \sinh ^{2}\theta _{n}\right. + \nonumber \\
&&\left. -\left( \widetilde{A}_{n}^{\mu }\widetilde{A}_{n}^{\mu \dagger }+
\widetilde{B}_{n}^{\mu }\widetilde{B}_{n}^{\mu \dagger }\right) \log \cosh
^{2}\theta _{n}\right] 
\label{entroop2}
\end{eqnarray}
for the tilde-string, respectively. Using the $G$-operators algebra
it is easy to show that
\begin{equation}
\left[ K-\widetilde{K}, G^\a \right] =0~~~,
~~~\left[ K-\widetilde{K}, G^\b \right] =0,
\label{ko}
\end{equation}
where the operators $G^\a$ and $G^\b$ are defined as
\be
G^\a = \sum\limits_n G^{\a}_{n}~~~,~~~
G^\b = \sum\limits_n G^{\b}_{n}.
\label{goptot}
\ee
In order to compute the entropy, one has to find the expectation value of
the operator (\ref{entroop1}) in the state (\ref{solT1}). 
In the thermo field dynamics it is postulated that
the physical properties of the system should be given in terms of operators
without tilde \cite{tu}. Therefore, we do not compute the expectation
value of $\widetilde{K}$ operator. 

It is useful to write the expectation value of the entropy operator in the
following form
\begin{eqnarray}
_{1} \! \left\langle \! \left\langle B\left( \beta_{T} \right) \right| \right.
K\left.\left| B\left( \beta_{T} \right) \right\rangle \! \right\rangle _{1}
&=&\,_{1} \! \left\langle \! \left\langle 0\left( \beta_{T} \right) \right| \right.
e^{F^{\dagger }}Ke^{F}\left. \left| 0\left( \beta_{T} \right) \right\rangle \!
\right\rangle _{1}  \nonumber \\
&=&_{1} \! \left\langle \! \left\langle 0\left( \beta_{T} \right) \right| \right.
e^{F^{\dagger }}e^{F}C\left. \left| 0\left( \beta_{T} \right) \right\rangle \!
\right\rangle _{1} +  \nonumber \\
&&+_{1} \! \left\langle \! \left\langle 0\left( \beta_{T} \right) \right| \right.
e^{F^{\dagger }}e^{F}K\left. \left| 0\left( \beta_{T} \right) \right\rangle \!
\right\rangle _{1},
\label{expvalentro}
\end{eqnarray}
where we are using the following notation 
\begin{eqnarray}
F =F\left( \beta _{T}\right)
=\exp \left[ -\sum\limits_{n=1}^{\infty }A_{n}^{\dagger }\left( \beta
_{T}\right) \cdot M\cdot B_{n}^{\dagger }\left( \beta _{T}\right) \right]
\label{foperat}
\end{eqnarray}
and
\begin{equation}
C=\left[ K,F\right]. 
\label{commkf}
\end{equation}
To compute the terms in (\ref{expvalentro}) we should work with the 
operators defined either at finite temperature or at zero temperature
since the action of these operators on the corresponding vacua is
known. The two ways of doing computations are completely equivalent.
The following relations are useful
\begin{eqnarray}
u_{m}^{2}A_{m}^{\rho \dagger }\left( \beta_{T} \right) A_{m}^{\rho }
\left(\beta \right) \left. \left| 0\left( \beta_{T} \right)
\right\rangle \! \right\rangle  &=&0, \nonumber\\
u_{m}v_{m}A_{m}^{\rho \dagger }\left( \beta_{T} \right) \widetilde{A}
_{m}^{\rho \dagger }\left( \beta_{T} \right) \left. \left| 0
\left( \beta_{T}\right) \right\rangle \! \right\rangle  &=&u_{m}v_{m}\left. 
\left| \left(1_{m}\right) ^{\rho \tilde{\r}};0;\beta_{T} \right\rangle \!
\right\rangle, 
\nonumber \\
u_{m}v_{m}\widetilde{A}_{m}^{\rho }\left( \beta_{T} \right) A_{m}^{\rho
}\left( \beta_{T} \right) \left. \left| 0\left( \beta_{T} \right) \right\rangle
\! \right\rangle  &=&0, \nonumber \\
v_{m}^{2}\widetilde{A}_{m}^{\rho }\left( \beta_{T} \right) \widetilde{A}
_{m}^{\rho \dagger }\left( \beta_{T} \right) \left. \left| 0\left( \beta
\right) \right\rangle \! \right\rangle  &=&v_{m}^{2}\left. \left| 0
\left( \beta_{T}\right) \right\rangle \! \right\rangle,
\label{zerofiniterel}
\end{eqnarray}
where we used the following shorthand notation 
\begin{equation}
\left. \left| \left( 1_{m}\right) ^{\rho \tilde{\rho}};0;\beta_{T}\right\rangle
\! \right\rangle =\left. \left| 1,\tilde{1};m,\widetilde{m};\beta_{T}
\right\rangle \! \right\rangle _{\alpha }^{\rho \widetilde{\rho }}\left. \left|
0\left( \beta_{T} \right) \right\rangle \! \right\rangle _{\beta },
\label{short1}
\end{equation}
and
\begin{equation}
\left. \left| 1,\widetilde{1};m,\widetilde{m};\beta_{T} \right\rangle
\! \right\rangle _{\alpha }^{\rho \widetilde{\rho }}=\left. \left|
0,...,1,...,0;0,...,\widetilde{1},...,0;...,m,...;...,\widetilde{m}
,...;\beta_{T} \right\rangle \! \right\rangle _{\alpha }^{\rho
\widetilde{\rho }}.
\label{short2}
\end{equation}
Here, the indices $\r$ and $\tilde{\r}$ indicate that the state 
has one quanta in
the $X^\r$ direction of space-time. The space-time is the same for the string 
as well as for the tilde-string since doubling the system does not mean doubling
the space-time. The tilde over the index just indicates that there is a state,
for the tilde-string on that direction, that should be taken into account. 
The first $1$ on the r.h.s. of (\ref{short2}) means that the quanta is
of string type, in the $\r$-th direction. Zero stands for the other directions.
The $\tilde{1}$ means that one quanta of tilde-string exists in that state and
in the same direction $\r$ and there are no other quanta in the other 
directions of space-time. $m$ is the mode of the string quanta, while 
$\tilde{m}$ is the mode of the tilde-string quanta. The modes that are on the
right-moving sectors of the closed string and tilde-string are denoted by
the index $\a$. Since the oscillators are independent in all directions and
for all different modes, the orthogonality relations among these states are
easy to write down. We also assume that the states are orthonormal.  

Using the properties of $G$-operators and the relations (\ref{zerofiniterel}) we can 
write the action of the entropy operator on the boundary state at finite 
temperature  under the form
\begin{eqnarray}
Ke^{F}\left. \left| 0\left( \beta_{T} \right) \right\rangle \! \right\rangle _{1}
&=&-\sum_{\mu ,\nu =1}^{24}\sum_{k=1}^{\infty }\frac{\left( -\right) ^{k}}{k!
}\left[ A_{n}^{\mu \dagger }\left( \beta_{T} \right) M^{\mu \nu }B_{n}^{\nu
\dagger }\left( \beta_{T} \right) \right] ^{k}\times   \nonumber \\
&&\times \sum_{m=1}^{\infty }\left\{ \log \tanh ^{2}\theta _{m}\sum_{\rho
=1}^{24}\left[ u_{m}v_{m}\left. \left| \left( 1_{m}\right) ^{\rho };0;\beta_{T}
\right\rangle \! \right\rangle \right] \right.   \nonumber \\
&&\left. +48\left[ v_{m}^{2}\log \tanh ^{2}\theta _{m}-\log \cosh
^{2}\theta _{m}\right] \left. \left| 0\left( \beta_{T} \right) \right\rangle
\! \right\rangle \right\}. 
\label{entroponD1}
\end{eqnarray}
The first term can be written in terms of the states (\ref{short2}) as
\begin{eqnarray}
&&\sum_{k=1}^{\infty }\frac{\left( -\right) ^{k}}{k!}\left[ A_{n}^{\mu
\dagger }\left( \beta_{T} \right) \right] ^{k}\left[ M_{\mu \nu }\right]
^{k}\left[ B_{n}^{\nu \dagger }\left( \beta_{T} \right) \right] ^{k}\times  
\nonumber \\
&&\times \left( -\right) \sum_{m=1}^{\infty }\left\{ \log \tanh ^{2}\theta
_{m}\sum_{\rho =1}^{24}\left[ u_{m}v_{m}\left. \left| \left( 1_{m}\right)
^{\rho };0;\beta_{T} \right\rangle \! \right\rangle \right] \right\}  
\nonumber \\
&=&\left. \left| 0,...,1,...,k,...,0;0,...,\widetilde{1}
,...,0;...,m,...,n,...;...,\widetilde{m},...;\beta_{T} \right\rangle
\! \right\rangle _{\alpha }^{\rho \widetilde{\rho }\mu }\otimes   \nonumber \\
&&\otimes \left. \left|
0,...,0,...,k,...,0;0,...,0,...,0;...,n,...;...;\beta_{T} \right\rangle
\! \right\rangle _{\beta }^{\nu }.
\label{entroponD2}
\end{eqnarray}
The relations (\ref{entroponD1}) and (\ref{entroponD2}) are important for 
computing the second term in (\ref{expvalentro}) while the first term can be
shown to be of the following form
\begin{eqnarray}
_{1} \! \left\langle \! \left\langle 0\left( \beta _{T}\right) \right| \right.
e^{F^{\dagger }}e^{F}C\left. \left| 0\left( \beta _{T}\right) \right\rangle
\! \right\rangle _{1} &=&2\prod_{m=1}^{\infty }\prod_{\mu =1}^{24}\prod_{\nu
=1}^{24}\sum_{k=0}^{\infty }
u_{m}^{2}\left( \beta_{T} \right)
\frac{\left( -\right) ^{2k+1}M_{\mu \nu }^{2k+2}%
}{k!\left( k+1\right) !}
\label{firsttermT}
\end{eqnarray}
After some tedious algebra one obtains from (\ref{entroponD2}) and
(\ref{firsttermT}) the following value for the entropy of the closed
string in the boundary state at finite temperature
\begin{eqnarray}
K_{Dp} = _{1} \! \left\langle \! \left\langle 
B\left( \beta _{T}\right) \right| \right.
K\left. \left| B\left( \beta _{T}\right) \right\rangle \! \right\rangle _{1}
&=&48\sum_{m=1}^{\infty }\left[\log \sinh ^{2}\theta _{m}
-{\sinh}^{2}\theta_m \log \tanh^{2}\theta _{m}\right] \nonumber \\
&&+2\prod_{m=1}^{\infty }\prod_{\mu=1}^{24}
\prod_{\nu =1}^{24}\sum_{k=0}^{\infty }{\cosh}^{2}\theta_m
\frac{\left( -\right)^{2k+2}M_{\mu \nu }^{2k+2}}{k!\left( k+1\right) !}.
\label{braneentropy}
\end{eqnarray}
The first two terms obtained above do not depend on the
gauge field $F_{ab}$. It is easy to see that in the $T\rightarrow 0$ limit 
the contribution of the oscillators to $K_{Dp}$ diverges. If we subtract the
infinity we are left with the last term that may converge but towards a 
positive number only. However, in the $ T\rightarrow \infty$, the contribution
of the oscillator behaves as $log(-1)$. This might be an indication that
the notion of temperature breaks down for arbitrary large temperature due to 
the similar phenomenon which occurs in string theory at Hagedorn temperature.
We do not have a good explanation for this behaviour. 

\section{Summary and discussions}

In this paper we constructed the bosonic boundary states at finite temperature 
which were obtained by solving the boundary conditions of the bosonic closed 
string. The boundary conditions at $T\neq 0$ were obtained by mapping the
corresponding boundary conditions at 
$T=0$ with thermal Bogoliubov transformations. Since the states at $T=0$ have
the interpretation of $Dp$-branes we interpret the solutions we have found as
$Dp$-branes at finite temperature. The construction was done in the framework
of the thermo field dynamics \cite{tu}. We reproduced the states obtained in a 
previous work \cite{ivv} given by (\ref{solT1}) and corrected some misprints 
in that paper and also, we obtained a new solution (\ref{solT2}) which 
represents the Bogoliubov map of the $Dp$-branes at zero-temperature. 

The crucial point in our construction was to map the boundary conditions at 
$T\neq 0$. We have adopted the point of view that the dependence on the 
temperature should be encoded in the operators which are in agreement with 
thermo field dynamics. We have obtained in this way two sets of boundary 
conditions, one for the original string and another one for its copy.
In \cite{ivv} another solution of one set of boundary conditions was discussed.
Since the total Fock space is a direct product of two copies of the Fock space
of a bosonic closed string, a linear combinations of boundary states 
described by the following operators can be given 
\footnote{These solutions were obtained in \cite{ivv}. There
the operators should be read at $T \neq 0$.}
\be
{\hat{O}}_{2,3}  =  
N_{p}\left( F,\beta_T \right)
\left(
\prod_{n=1}^{\infty}e^{-A^{\dagger}_n 
(\b_T ) \cdot M \cdot B^{\dagger}_n (\b_T )}
\times 1 
\pm 1 \times
\prod_{n=1}^{\infty}e^{-
{\tilde{A}}^{\dagger}_n (\b_T )\cdot M \cdot {\tilde{B}}^{\dagger}_n (\b_T )}
\right)
\label{opsolT3}
\ee
where an implicit delta function that localizes the center of mass of the
string and the tilde-string are assumed to exist. The operators
(\ref{opsolT3}) act on the thermal vacuum. We note that the vectors
constructed with (\ref{opsolT3}) satisfy one set of boundary conditions,
say, for the string, but do not satisfy the boundary conditions for
the tilde-string.

For the $Dp$-brane (\ref{solT1}) which has the same form as the $Dp$-brane
at $T = 0$, the entropy was given in (\ref{braneentropy}). We found that
in the limit $ T \rightarrow 0$ the contribution of the oscillators diverges.
If this is subtracted from the total value of the entropy what is left
is a series of positive terms  with a slower increment than the exponential
function.
When the limit $T \rightarrow \infty$ is taken, the oscillator contribution
breaks down since it involves the logarithm of negative unity. One possible
explanation is that using bosonic strings is inappropriat for describing
$Dp$-branes at high temperatures since the phase transition of string theory
around the Hagedorn temperature makes the statistics of strings 
unfeasible \cite{aw}, and consequently the boundary states.
Responsible for this phenomenon are the tachyons, the gravitons and 
the dilaton. In the present case only physical modes have been considered 
since we have worked in the light-cone gauge. Nevertheless, the contributions
of the massless modes can induce a bad behaviour for the entropy. It is
worthwhile to investigate the contributions that can be obtained from the
fermionic part in the case of super-branes. Another possibility would be that
$Dp$-branes at finite temperature are not acceptable boundary states for
closed string theory at $T\neq 0$.
However, to give a definitive answer to these problems, further 
investigations of the thermal properties of the boundary states 
should be performed. We hope to report on these topics in a future 
work \cite{cawv}.  
 
{\bf Acknowledgements}
We would like to thank N. Berkovits, P. K. Panda,
B. M. Pimentel, H. Q. Placido, D. L. Nedel and B. C. Vallilo for useful
discussions. I. V. V. also acknowledge to M. A. De Andrade and J. A.
Helayel-Neto for hospitality at GFT-UCP where part of this work was done.
This work was supported by FAPESP (I.V.V.) and CAPES (A.L.G.).

\vskip 0.5cm
{\bf Appendix}
\vskip 0.5cm

In this Appendix we review some basic relations of thermo field theory
with focus in the bosonic harmonic oscillator. We closely follow \cite{tu}. 

The basic idea of the Thermo Field Dynamics (TFD) is to
construct statistical averages of some operator $A$ in terms of expectations
values in a vacuum that presents a dependence with the temperature:
\begin{equation}
\left\langle A\right\rangle \equiv Z^{-1}\left( \beta _{T}\right)
Tr[e^{-\beta _{T}H}A]=\left\langle \! \left\langle 0\left( \beta _{T}\right) \right| A\left|
0\left( \beta _{T}\right) \right\rangle \! \right\rangle ,
\label{average}
\end{equation}
where $\beta _{T}=\left( k_{B}T\right) ^{-1}$ is the Botzmann's constant and.
\begin{equation}
H\left| n\right\rangle=E_{n}\left| n\right\rangle.
\label{hamilt}
\end{equation}
The eigenstates of the Hamiltonian $H$ of the system are supposed to be 
orthonormal. The partition function is defined by the usual relation
\begin{equation}
Z\left( \beta _{T}\right) =Tr[e^{-\beta _{T}H}]
\label{partfunct}.
\end{equation}
It is possible to realize such of idea if a duplication of the physical 
Hilbert space is performed, i.e. if one considers The Hilbert space as 
direct product of two spaces: the original one and a nonphysical copy of it. 
The non physical or auxiliary subspace is represented by a tilde and it is 
acted on by a copy of the operators of the original system, e. g.
\begin{equation}
\widetilde{H}\left| \widetilde{n}\right\rangle =E_{n}\left| \widetilde{n}
\right\rangle. 
\label{copyhamilt}
\end{equation}
As a consequence any state in the total Hilbert space has the following form
\begin{equation}
\left| n,\widetilde{m}\right\rangle =\left| n\right\rangle \otimes \left| 
\widetilde{m}\right\rangle.
\label{statetotal}
\end{equation}
The temperature dependent vacuum can be written as 
\begin{equation}
\left. \left| 0\left( \beta _{T}\right) \right\rangle \! \right\rangle =
Z^{-1/2}\left( \beta_{T}\right) \sum_{n}e^{-\frac{1}{2}\beta _{T}E_{n}}
\left| n,\widetilde{m}\right\rangle ,
\end{equation}
and it is called ``thermal vacuum''.

One simple system for which the above construction is straightforward is
the single bosonic oscillator
\begin{equation}
H=\omega a^{\dagger }a,
\label{hamiltosc}
\end{equation}
with the known algebra
\begin{equation}
\lbrack a^{\dagger },a]=1;\quad \left[ a,a\right] =
[a^{\dagger },a^{\dagger}]=0,
\label{algosc}
\end{equation}
and the Fock space spanned by the following states
\begin{equation}
\left| n\right\rangle =\frac{(a^{\dagger })^{n}}{\sqrt{n!}}\left|
0\right\rangle ,
\end{equation}
for all positive integers $n$. The tilde oscillator has identical properties
\begin{equation}
\widetilde{H}=\omega \widetilde{a}^{\dagger }\widetilde{a},
\label{tildehamilt}
\end{equation}
and
\begin{equation}
\lbrack \tilde{a}^{\dagger },\tilde{a}]=1;\quad \left[ \tilde{a},\tilde{a}
\right] =[\tilde{a}^{\dagger },\tilde{a}^{\dagger }]=0.
\end{equation}
It is independent of the first oscillator, i. e. the two Hilbert spaces 
are completely ortogonal:
\begin{equation}
\lbrack a,\tilde{a}]=\left[ a,\tilde{a}^{\dagger }\right] =[a^{\dagger },
\tilde{a}^{\dagger }]=[a^{\dagger },\tilde{a}]=0.
\end{equation}
The duplicated vacuum is defined by
\begin{equation}
a\left. \left| 0\right\rangle \! \right\rangle =\widetilde{a}\left. \left|
0\right\rangle \! \right\rangle ,
\label{duplvacosc}
\end{equation}
which has a solution in terms of the two vacua 
\begin{equation}
\left. \left| 0\right\rangle \! \right\rangle \equiv \left| 0,\widetilde{0}
\right\rangle =\left| 0\right\rangle \otimes \widetilde{\left|
0\right\rangle }.
\end{equation}
The Hamiltonian of the total system is defined by the following relation
\begin{equation}
\widehat{H}=H-\widetilde{H}.
\end{equation}
A simple algebra shows that the thermal vacuum can written in terms of
a coherent state
\begin{equation}
\left. \left| 0\left( \beta _{T}\right) \right\rangle \! \right\rangle 
=\frac{1}{u\left( \beta _{T}\right) }\exp \left[ \frac{v\left( \beta
_{T}\right) }{u\left( \beta _{T}\right) }a^{\dagger }\tilde{a}^{\dagger
}\right] \left. \left| 0\right\rangle \! \right\rangle .
\label{vacthermcoh}
\end{equation}
The partition function of the oscillator is given by
\begin{equation}
Z\left( \beta _{T}\right) =\frac{1}{1-e^{-\beta _{T}\omega }},
\label{partosc}
\end{equation}
ensuring that the thermal vacuum has unit norm. Also, the 
temperature dependent coefficients are defined by the relations
\begin{eqnarray}
u\left( \beta _{T}\right)  &\equiv &\left( 1-e^{-\beta _{T}\omega }\right)
^{-1/2}= \cosh \theta \left( \beta _{T}\right) , \\
v\left( \beta _{T}\right)  &\equiv &\left( e^{\beta _{T}\omega }-1\right)
^{-1/2} = \sinh \theta \left( \beta _{T}\right)   .
\label{tempcoeff}
\end{eqnarray}
If we consider  unitary Bogoliubov operators having the form
\begin{equation}
G_{B}\equiv -i\theta \left( \beta _{T}\right) (a\tilde{a}-a^{\dagger }
\tilde{a}^{\dagger }),
\label{bogooposc}
\end{equation}
it is easy to show that the thermal vacuum is obtained from the total vacuum at
$T = 0$ by a simple Bogoliubov transformation
\begin{equation}
\left. \left| 0\left( \beta _{T}\right) \right\rangle \! \right\rangle =e^{-iG_{B}}\left. \left|
0\right\rangle \! \right\rangle .
\label{totalvactosc}
\end{equation}
The operators at finite temperature are also generated by the Bogoliubov 
operators
\begin{eqnarray}
a\left( \beta_{T} \right)  &=&e^{-iG_{b}}a\;e^{iG_{b}}=u\left( \beta _{T}
\right)
a-v\left( \beta _{T}\right) \tilde{a}^{\dagger }, \\
\tilde{a}\left( \beta_{T} \right)  &=&e^{-iG_{b}}\tilde{a}\;e^{iG_{b}}=u\left(
\beta _{T}\right) \tilde{a}-v\left( \beta _{T}\right) a^{\dagger },
\end{eqnarray}
and they act on the thermal vacuum (justifying its name) as
\begin{equation}
a\left( \beta_{T} \right) \left. \left| 0\left( \beta _{T}\right) 
\right\rangle \! \right\rangle = \tilde{a}\left( \beta_{T} \right)
\left. \left| 0\left( \beta _{T}\right) \right\rangle \! \right\rangle =0.
\end{equation}
With this construction, the Fock space is spanned by the vectors
\begin{eqnarray}
&&\left. \left| 0\left( \beta _{T}\right) \right\rangle \! \right\rangle ,\quad a^{\dagger }\left(
\beta _{T}\right) \left. \left| 0\left( \beta _{T}\right) \right\rangle \! \right\rangle ,\quad 
\tilde{a}^{\dagger }\left( \beta _{T}\right) \left. \left| 0\left( \beta
_{T}\right) \right\rangle \! \right\rangle ,...  \nonumber \\
&&\frac{1}{\sqrt{n!}}\frac{1}{\sqrt{m!}}[a^{\dagger }\left( \beta
_{T}\right) ]^{n}[\tilde{a}^{\dagger }\left( \beta _{T}\right) ]^{m}
\left. \left|0\left( \beta _{T}\right) \right\rangle \! \right\rangle ,...
\end{eqnarray}
The commutation relation between the temperature dependent operators are the 
same as for the duplicated system at zero temperature. It is easy to show that
the $G$-operator is a constant of motion:
\begin{eqnarray}
\lbrack G,\widehat{H}]=[G,H-\widetilde{H}]=0.
\end{eqnarray}

The construction presented above can be straightforward extended to set of
infinite free bosonic oscillators. In that case the Bogoliubov operator is
given by
\begin{eqnarray}
G_{B}\equiv -i\sum_{n}\theta \left( \beta _{T}\right) (a_{n}\tilde{a}
_{n}-a_{n}^{\dagger }\tilde{a}_{n}^{\dagger }),
\end{eqnarray}
and the thermal vacuum can be expressed as
\begin{eqnarray}
\left. \left| 0\left( \beta _{T}\right) \right\rangle \! \right\rangle =e^{-\frac{1}{2}K}
\exp\{\sum_{n}a_{n}^{\dagger }\tilde{a}_{n}^{\dagger }\}\left. \left|
0\right\rangle \! \right\rangle,
\end{eqnarray}
where
\begin{eqnarray}
K=-\sum_{n}\{a_{n}^{\dagger }a_{n}\log \sinh ^{2}\theta
_{n}-a_{n}a_{n}^{\dagger }\log \cosh ^{2}\theta _{n}\}.
\end{eqnarray}
$K$ is called entropy operator. This interpretation comes from the fact that the
vacuum expectation value of $K$ times the Boltzmann constant $k_{B}$ is the
entropy of the physical system.

\end{document}